\documentclass[useAMS,usenatbib]{mn2e}
\usepackage{graphicx}
\usepackage{epstopdf}
\usepackage{amsmath}
\usepackage{amssymb}

\title[Accretion at Large Radius]{The Effect of Accretion Environment at Large Radius on Hot Accretion Flows}
\author[Yang \& Bu]{Xiao-Hong Yang$^1$\thanks{E-mail: yangxh@cqu.edu.cn (XH)} and De-Fu Bu$^2$\\
$^1$ Department of Physics, Chongqing University, Chongqing 400044, China\\
$^{2}$Key Laboratory for Research in Galaxies and Cosmology, Shanghai Astronomical Observatory, Chinese Academy of Sciences,\\ 80 Nandan Road, Shanghai 200030, China}

\begin{document}


\pagerange{\pageref{firstpage}--\pageref{lastpage}} \pubyear{20**}

\maketitle

\label{firstpage}

\begin{abstract}
We study the effects of accretion environment (gas density, temperature, and angular momentum) at large radii ($\sim 10$pc) on luminosity of hot accretion flows. The radiative feedback effects from the accretion flow on the accretion environment are also self-consistently taken into account. We find that the slowly rotating flows at large radii can significantly deviate from Bondi accretion when radiation heating and cooling are considered. We further find that when the temperature of environment gas is low (e.g. $T=2\times 10^7$K), the luminosity of hot accretion flows is high. When the temperature of gas is high (e.g. $T\geq4\times 10^7$K), the luminosity of hot accretion flow significantly deceases. The environment gas density can also significantly influence the luminosity of accretion flows. When density is higher than $\sim 4\times 10^{-22}\text{g} \text{cm}^{-3}$ and temperature is lower than $2\times 10^7$K, hot accretion flow with luminosity lower than $2\%L_{\text{Edd}}$ is not present. Therefore, the parsec-scale environment density and temperature are two important parameters to determine the luminosity. The results are also useful for the subgrid models adopted by the cosmological simulations.
\end{abstract}

\begin{keywords}
accretion, accretion discs--black hole physics---hydrodynamics.
\end{keywords}

\section{Introduction}
Mass accretion rate is a key parameter to determine the properties of black hole accretion flows. According to the temperature of accretion flow, black hole accretion models can be divided into two series, i.e. cold and hot accretion flows. The cold accretion flow models include standard thin disc solution (Shakura \& Sunyaev 1973) and slim disc solution (Abramowicz et al. 1988). Standard thin disc solution exists when the mass accretion rate is lower than the Eddington rate, i.e. $\dot{M}\lesssim \dot{M}_{\text{Edd}} (\equiv10 L_{\text{Edd}}/c^2)$; slim disc model is present when $\dot{M}\gtrsim \dot{M}_{\text{Edd}}$. For the hot accretion flows, when $\dot{M} \lesssim \alpha^2\dot{M}_{\text{Edd}}$, where $\alpha$ is the viscous parameter, the solution is described by the advection-dominated accretion flows (ADAFs; Narayan \& Yi 1994; 1995; Narayan \& McClintock 2008; Yuan \& Narayan 2014); when $\dot{M}\gtrsim\alpha^2\dot{M}_{\text{Edd}}$, the solution is described by luminous hot accretion flow (LHAFs; Yuan 2001; 2003).

An important question is that what determines the mass accretion rate? The mass accretion rate can be determined by two factors. The first one is the environment (e.g. gas temperature, density, angular momentum) of the gas supply which directly determines the boundary conditions of the accretion flow. The other one is the active galactic nuclei (AGNs) feedback effects (e.g. momentum feedback and energy feedback) which affect the mass accretion rate by affecting the mass supply environment (e.g. Ciotti \& Ostriker 1997, 2001, 2007; Ciotti, Ostriker \& Proga 2009; Ostriker et al. 2010). The two factors are interrelated. The remarkable correlations between host galaxy properties and the mass of the supermassive black holes (e.g. Magorrian et al. 1998; Ferrarese \& Merritt 2000; Gebhardt et al. 2000; Kormendy \& Bender 2009) provide a clue to understand the two factors. Currently, it is impossible to directly simulate the region from the black hole to the host galaxy scale because the dynamical range is too huge (many orders of magnitude in radial direction). Therefore, there are still uncertainties about the details of feedback process. Some authors focus on AGNs feedback effects at parsec scale (e.g. Kurosawa \& Proga 2009; Yuan \& Li 2011; Liu et al. 2013). There are also authors focus on AGNs feedback effects on kilo parsec scale (e.g. Di Matteo et al. 2005; Ciotti, Ostriker \& Proga 2009; Gan et al. 2014; Ciotti et al. 2017). The study of feedback effects at parsec scale is more direct to study interaction between the activities of AGNs and the accretion environment, because in these studies the inner boundary of computational domain can be placed at the location close to the boundary of accretion disk.

Cold and hot accretion flow models have different applications and feedback effects. The cold standard thin disk is often used to understand the properties of quasar, while the hot accretion flow is often used to explain the properties of low-luminosity AGNs (LLAGNs) (Pellegrini 2005; Soria et al. 2006). LLAGNs and quasars emit different spectra, which causes different radiation feedback effects. For a quasar, majority of the emitted photons are UV photons which can drive parsec-scale outflows by line force (e.g. Proga, Stone, \& Kallman 2000; Proga 2007; Proga, Ostriker, \& Kurosawa 2008; Kurosawa \& Proga 2009). For LLAGNs, a great number of high-energy photons from the innermost region of accretion flows can heat the electrons at the Bondi radii. If the radiation is strong enough, e.g. $L\ga2\%L_{\text{Edd}}$, the temperature of the electrons at the Bondi radius will be heated to be above the virial temperature, so that the accretion process will be stopped and the intermittent activity will take place. Such radiation feedback mechanism was proposed by Yuan \& Li (2011) as the origin of the intermittent activity of some compact young radio sources with $L\ga 2\%L_{\rm Edd}$.

In this paper, we will study the effects of accretion environment(e.g. gas temperature, density, and angular momentum) on LLAGNs. The radiative feedback effects from the accretion flow on the accretion environment are also self-consistently taken into account. The motivations are as follows. 1) LLAGNs universally exist. For example, Galactic Centre, i.e. Sgr A$^*$, is a LLAGNs with $3\times10^{-9}L_{\text{Edd}}$. Then a question is that what kind of environment can harbour a LLAGN. 2) Gan et al. (2014) studied the effect of strong radiative feedback from the LLAGNs, but they focus on the region of 2.5 pc $-$ 250 kpc. Yuan et al. (2009) pointed out that in the parsec and sub-parsec region, the global Compton scattering effect is already important. However, the effects of feedback from LLAGNs on parsec and sub-parsec scale have not been studied. The sub-pc and pc region covers the region from the boundary of LLAGNs to Bondi radius. 3) In the cosmological simulations studying galaxy formation and evolution (e.g. Springel et al. 2005; Teyssier et al. 2011; Vogelsberger et al. 2014; Khandai et al. 2015; Negri \& Volonteri 2017), the Bondi radius can at most be marginally resolved. The authors use sub-grid models to study the black hole growth and AGNs feedback. The most commonly used sub-grid model is the Bondi accretion model (e.g. Springel et al. 2005). However, the Bondi accretion model is only for adiabatic, spherical accretion. Thus, the accretion rate estimated by the Bondi formula may be not correct. The results in this paper can be adopted by sub-grid models in cosmological simulations

The paper is organized as follows. In section 2, we describe our models and method; In section 3, we present our results; Section 4 is devoted to conclusions and discussions.

\section{Model and Method}

The hot accretion flow around a black hole emits high-energy photons
(e.g. X-ray) to heat the flow at sub-pc and pc scale. The irradiated
flow is the gas source that fuels the hot accretion flow. Thus,
there is strong interaction between the hot accretion flow and the
irradiated flow. Here, we focus on the sub-pc and pc-scale flows
irradiated by hot accretion flows. The accretion luminosity of hot
accretion flows is self-consistently determined based on the mass
accretion rate through the inner boundary. Most of the radiation is
from the hot accretion flows inside $20r_{\text{s}}$ ($r_{\text{s}}$
is Schwarzschild radius) and tends to be isotropic. We set that the
inner boundary of the computation domain is much larger than the
radius ($20r_{\text{s}}$) inside which most of the radiation is
produced by hot accretion flows. Therefore, the radiation of hot
accretion flows can be approximated to be from a point object
located at $r=0$ and be isotropic.

\begin{table}
\caption[]{Summary of models}
\label{}\tiny
\begin{center}
\begin{tabular}{ccccc}
\hline\noalign{\smallskip} \hline\noalign{\smallskip}

Run &  $T_{\text{X}}$ & $\rho_0$                    & $T_0$     & $r_{\text{cir}}$  \\
    &  ($10^8$K)      & ($10^{-22}\text{g cm}^{-3}$)& ($10^7$K) & ($r_{\text{s}})$  \\
(1) & (2)             & (3)                         &  (4)      &     (5)           \\

\hline\noalign{\smallskip}
1a   & 1 &1       &   2   & 350  \\
2a   & 1 &1       &   4   & 350  \\
3a   & 1 &1       &   6   & 350  \\
4a   & 1 &1       &   10  & 350  \\

5a   & 1 &2       &   2   & 350  \\
6a   & 1 &2       &   4   & 350  \\
7a   & 1 &2       &   6   & 350  \\
8a   & 1 &2       &   10   & 350 \\

9a   & 1 &3       &   2   & 350  \\
10a   & 1 &3       &   4   & 350 \\
11a   & 1 &3       &   6   & 350 \\
12a   & 1 &3       &   10   & 350 \\

13a   & 1 &4       &   2   & 350 \\
14a   & 1 &4       &   4   & 350 \\
\hline\noalign{\smallskip}
1b   & 1  &1       &   2   & 150  \\
2b   & 1  &1       &   4   & 150  \\
3b   & 1  &1       &   6   & 150  \\
\hline\noalign{\smallskip}
1c   & 10 &1       &   2   & 350  \\
2c   & 10 &1       &   4   & 350  \\
3c   & 10 &1       &   6   & 350  \\

 \hline\noalign{\smallskip}
\end{tabular}
\end{center}

\begin{list}{}
\item\scriptsize{\textit{Note}. Column (1): the number of our models. Column (2) the Compton temperature of X-ray radiation from hot accretion flows. Columns (3), (4) and 5: the density, temperature, and `circularization radius' of the gas at the outer
boundary, respectively.}
\end{list}
\end{table}

\subsection{Method}
For numerically simulating the irradiated flows between an inner
radius $r_{\text{in}}$ and an outer radius $r_{\text{out}}$, we use
the ZEUS-MP code (Hayes et al. 2006) to solve the following set of
HD equations:
\begin{equation}
\frac{d\rho}{dt}+\rho\nabla\cdot \mathbf{v}=0,
\end{equation}

\begin{equation}
\rho\frac{d\mathbf{v}}{dt}=-\nabla p-\rho\nabla \psi+\rho
\mathbf{F}_{\text{rad}},
\end{equation}

\begin{equation}
\rho\frac{d(e/\rho)}{dt}=-p\nabla\cdot\mathbf{v}+\dot{E},
\end{equation}
where $\rho$, $p$, $\mathbf{v}$, $e$, and $\psi$ are mass density, gas pressure, velocity, internal energy, and gravitational potential, respectively. We adopt an equation of state of ideal gas $p=(\gamma -1)e$ with the adiabatic index $\gamma =5/3$ and employ the pseudo-Newtonian potential, $\psi=-GM/(r-r_{\text{s}})$ (Paczy\'{n}sky \& Wiita 1980), where $M$ and $G$ are the centre black hole mass and the gravitational constant, respectively, and $r_{\text{s}}\equiv2GM/c^2$. In Equation (2), $\mathbf{F}_{\text{rad}}$ is the radiation force per unit mass due to Compton scattering and is given by $\rho\mathbf{F}_{\text{rad}}=\frac{\chi}{c}\frac{L}{4\pi r^2}$, where $L$ is luminosity and $\chi$ ($=\rho \sigma_{\text{T}}/m_{\text{p}}$) is the Compton scattering opacity. In a hot accretion flow, its luminosity is much smaller than the Eddington luminosity so that $\rho\mathbf{F}_{\text{rad}}$ is not important.

In Equation (3), $\dot{E}$ is the net gas energy change rate due to heating and cooling. We consider Compton heating and cooling, bremsstrahlung cooling, photoionization heating, and line and recombination continuum cooling. Therefore, $\dot{E}$ depends on the gas temperature $T$, the Compton temperature $T_{\text{X}}$ characterizing the spectrum of radiation from the central source, and the photoionization parameter $\xi$.  A good approximation of $\dot{E}$, valid for gas temperature $T\geq 10^4 \text{K}$, is referred to Equation (A32) of Sazonov et al. (2005). The net energy change rate reads

\begin{equation}
\dot{E} =n^2(S_1+S_2+S_3),
\end{equation}
where $n$ is the number density of H nucleus,

\begin{equation}
S_1=-3.8\times10^{-27}T^{\frac{1}{2}}
\end{equation}
is the bremsstrahlung losses,

\begin{equation}
S_2=4.1\times10^{35}(T_{\rm X}-T)\xi
\end{equation}
is the Compton heating and cooling. $S_3$ is the sum of photoionization heating and line and recombination continuum cooling. For formula of $S_3$, we refer to Equation (A32) of Sazonov et al. (2005).

The Compton temperature ($T_{\text{X}}$) of the X-ray radiation depends on energy spectra of X-ray radiation from central source. For a quasar, $T_{\text{X}}=1.9\times10^7$ K (Sazonov et al. 2005). For a low-luminosity AGN, $T_{\text{X}}$ will be much higher (Yuan et al. 2009; Xie et al. 2017). Here, we set $T_{\text{X}}=10^8$ and $10^9 \text{K}$ for the comparison of effect of Compton temperature on feedback. The $\xi$ to determine the gas photoionization state is given by
\begin{equation}
\xi\equiv\frac{L}{nr^2}e^{-\tau_{\text{x}}},
\end{equation}
where $\tau_{\text{x}}$ ($=\int_0^r \rho \kappa_{\text{x}}dr$, where
$\kappa_{\text{x}}$ is the X-ray opacity) is the X-ray scattering
optical depth in the radial direction, $r$ is the distance from the
central source, $n=\rho/(\mu m_{\text{p}})$ is the number density of
the local gas located at $r$, and $\mu$ is the mean molecular
weight. We set $\mu=1$ and $\kappa_{\text{x}}=0.4
\text{cm}^2\text{g}^{-1}$ because Thomson scattering dominates the
attenuation.

\begin{figure*}
\scalebox{0.45}[0.45]{{\includegraphics[bb=26 5 504 380]{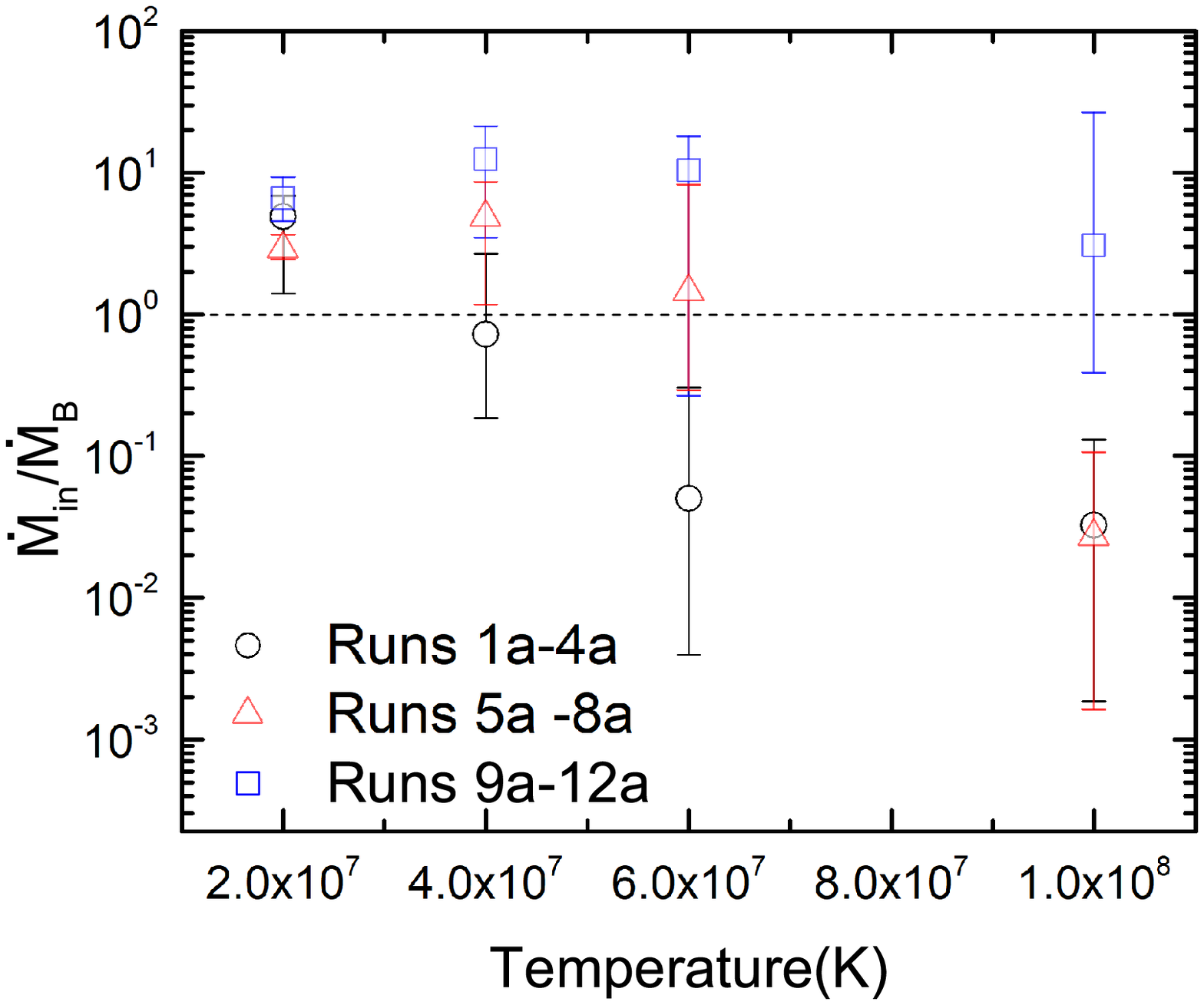}}}
\scalebox{0.45}[0.45]{{\includegraphics[bb=26 5 504 380]{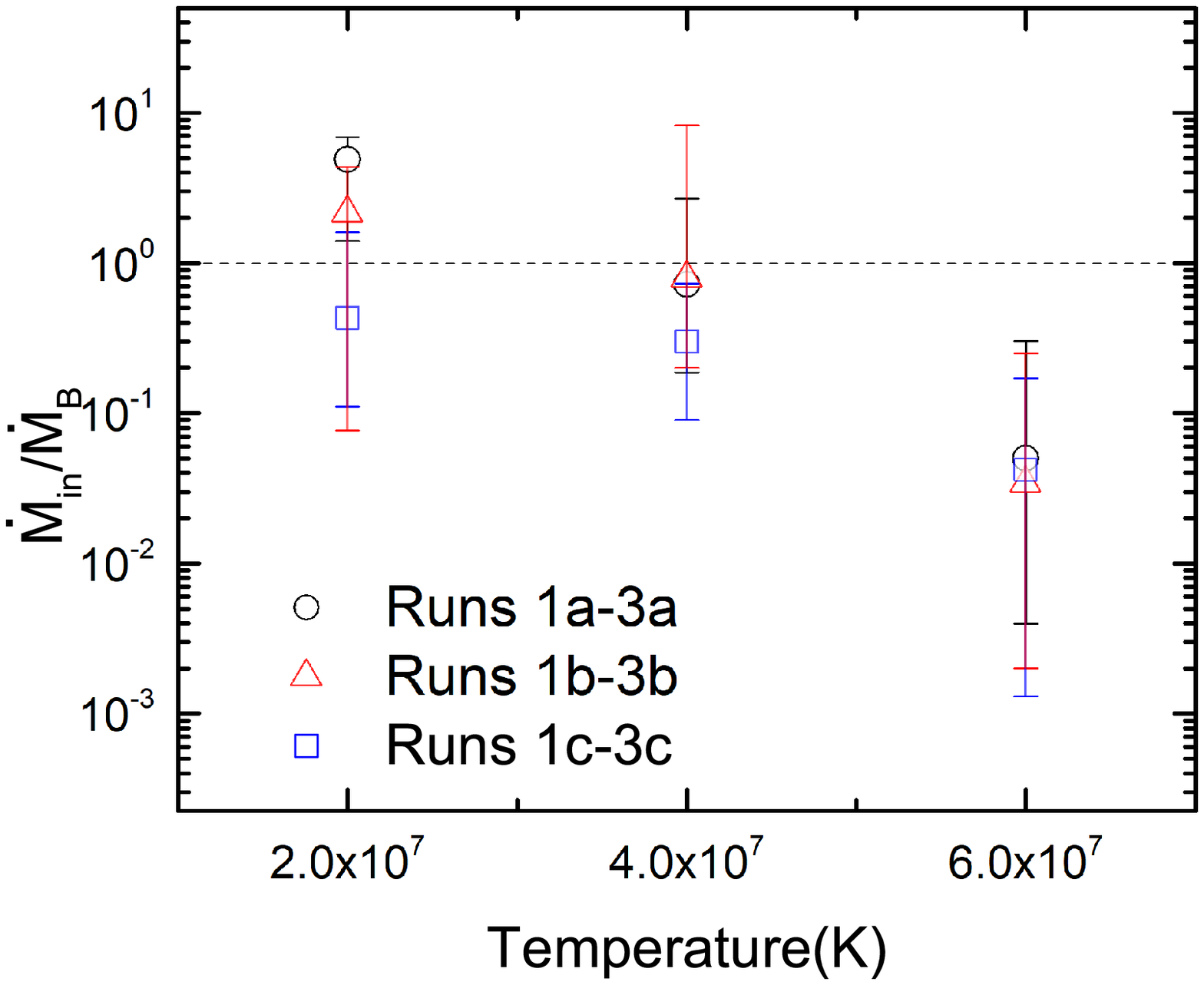}}}
\ \centering \caption{The mass inflow rate (in units of the Bondi accretion rate) through the inner boundary against the temperatures at the outer boundary. The left-hand panel is for runs 1a$-$14a and the right-hand panel is for runs 1a$-$3a, runs 1b$-$3b and runs 1c$-$3c. Symbols indicate the average values of luminosities, and error bars indicate the change range of accretion rate. } \label{fig 1}
\end{figure*}

\begin{figure*}
\scalebox{0.45}[0.45]{{\includegraphics[bb=26 5 504 380]{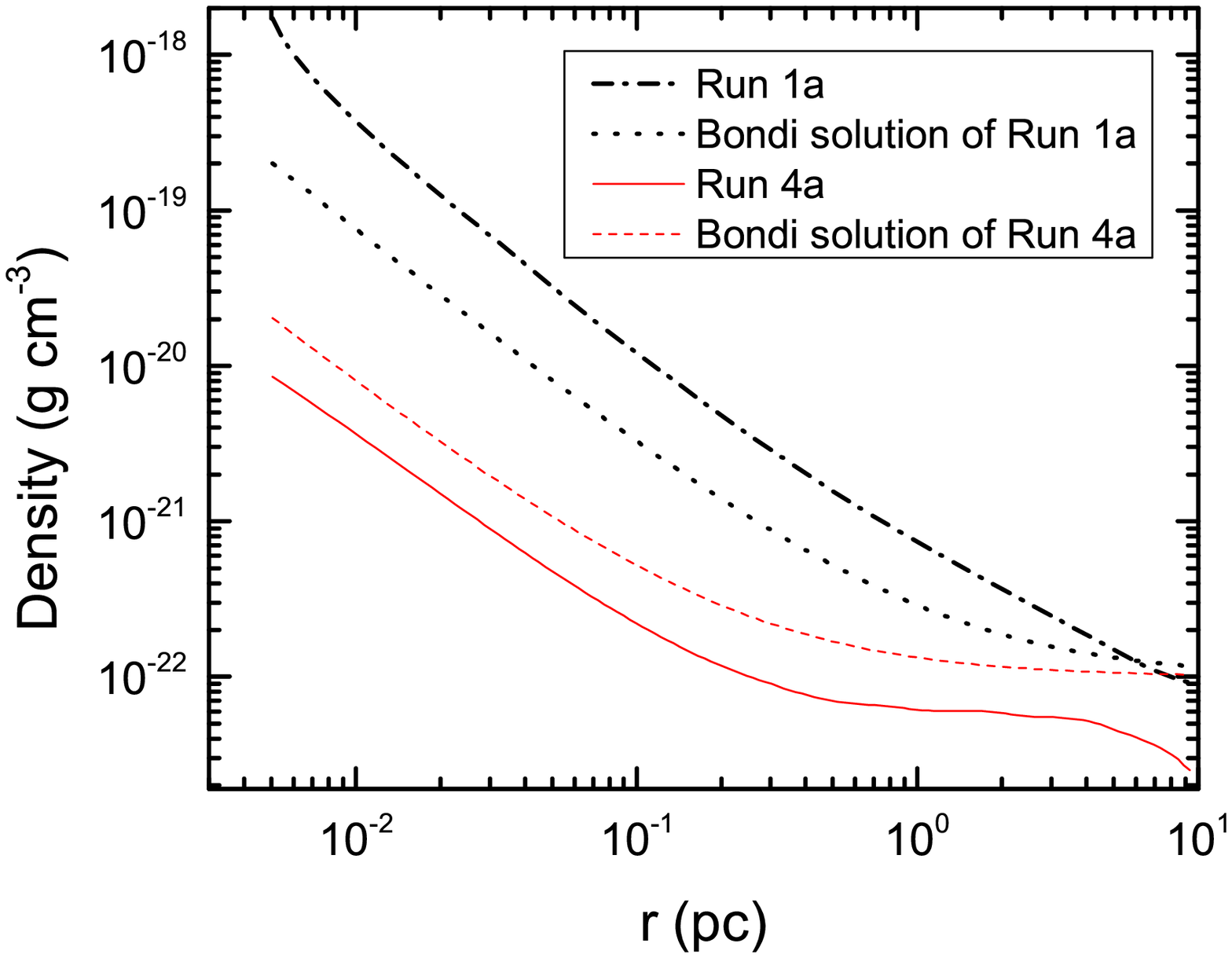}}}
\scalebox{0.45}[0.45]{{\includegraphics[bb=26 5 504 380]{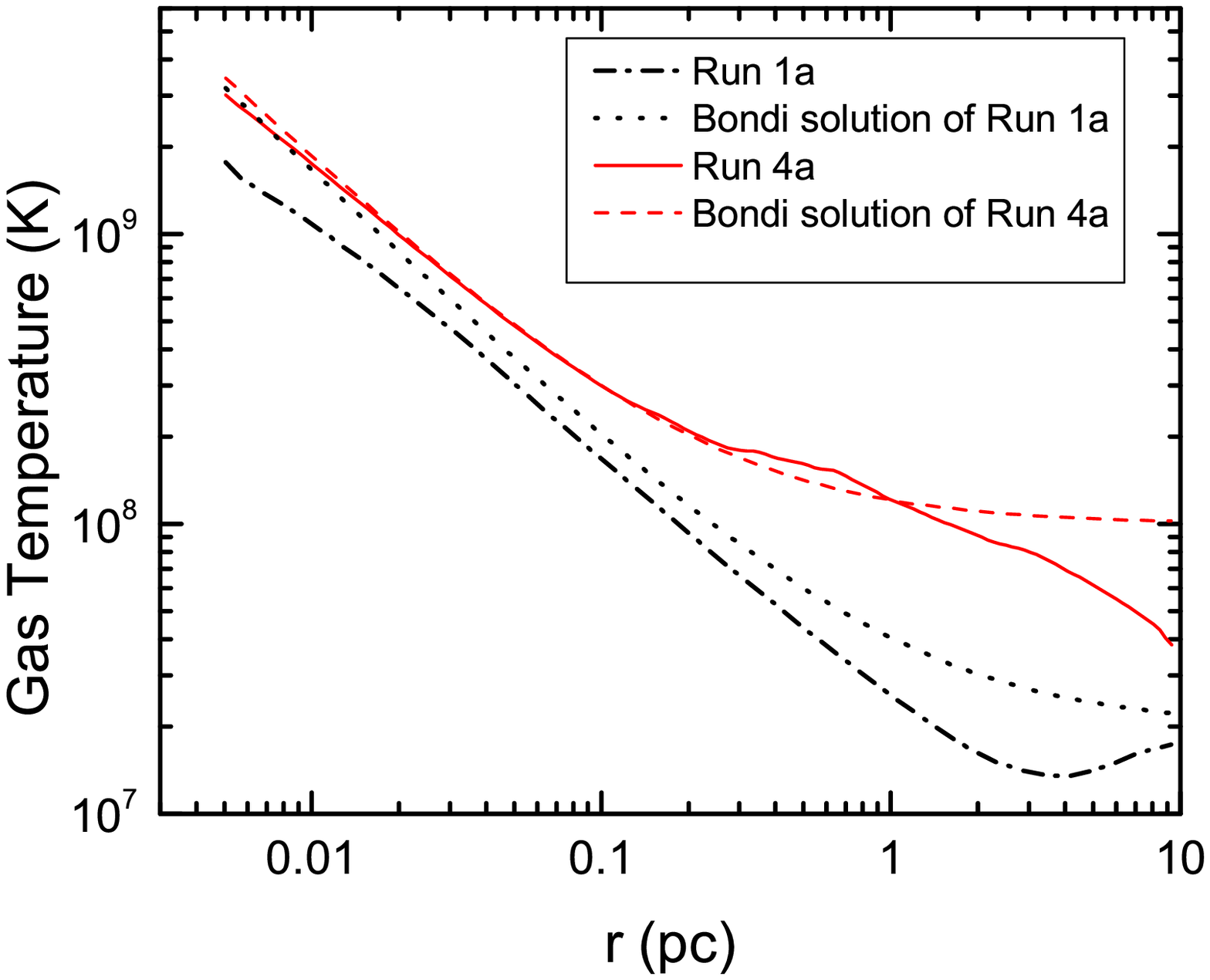}}}
\ \centering \caption{The radial profiles of time-averaged and angle-averaged density (left-hand panel) and gas temperature (right-hand panel) in runs 1a and 4a. } \label{fig 2}
\end{figure*}

\begin{figure*}
\scalebox{0.45}[0.45]{{\includegraphics[bb=26 5 504 380]{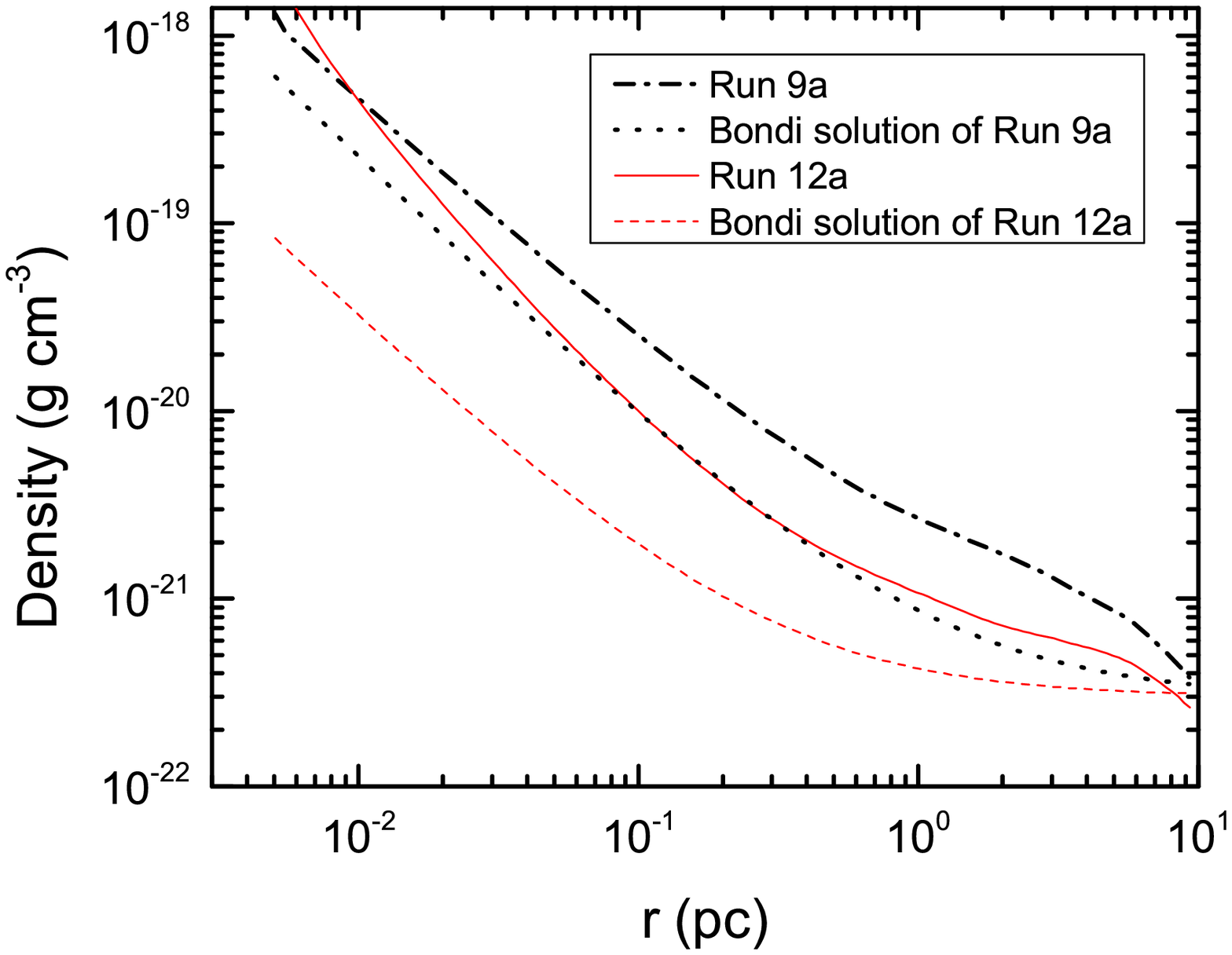}}}
\scalebox{0.45}[0.45]{{\includegraphics[bb=26 5 504 380]{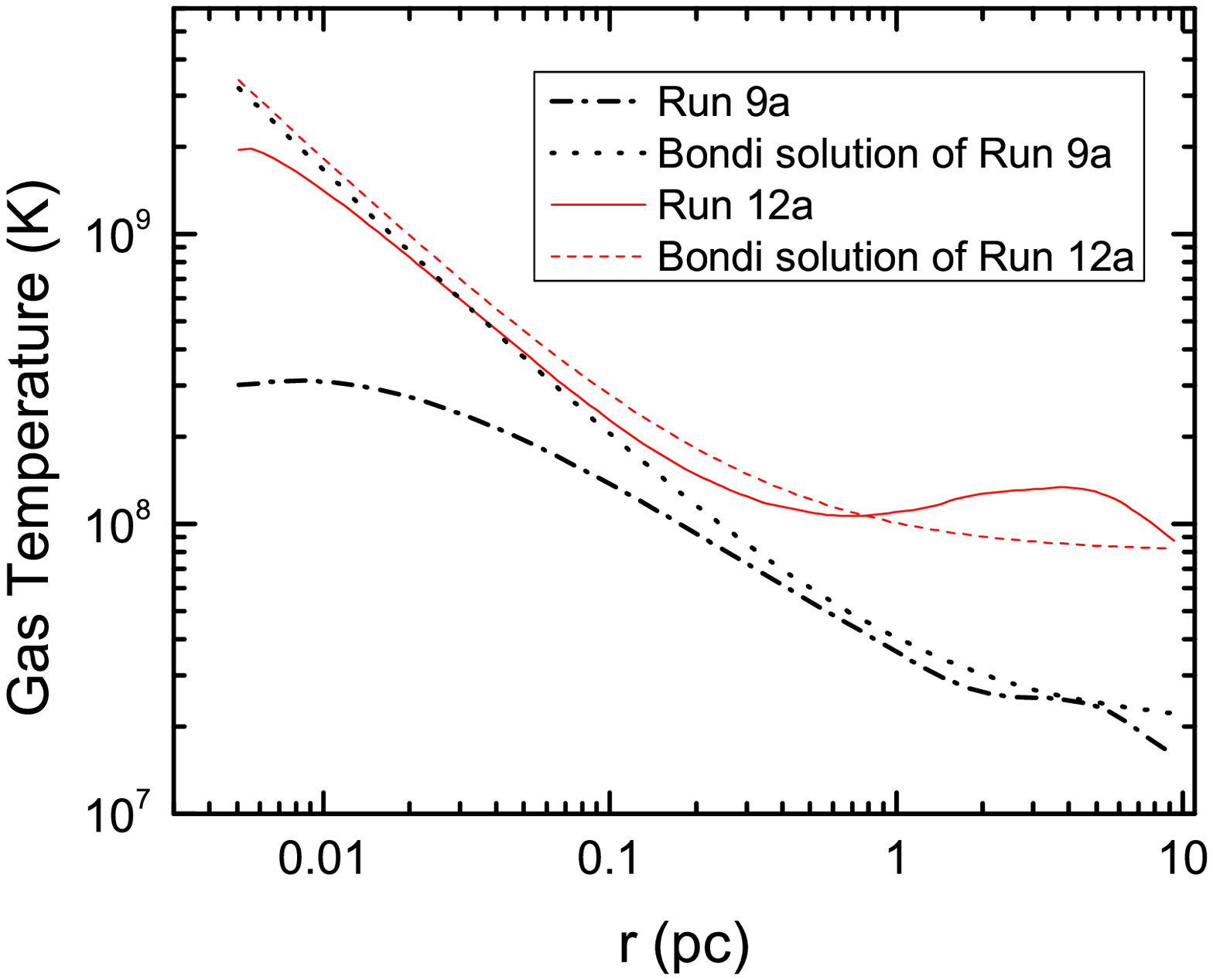}}}
\ \centering \caption{The radial profiles of time-averaged and angle-averaged density (left-hand panel) and gas temperature (right-hand panel) in runs 9a and 12a. } \label{fig 3}
\end{figure*}

\subsection{Coupling of luminosity and mass accretion rate}

The luminosity of a hot accretion flow depends on the accretion rate
and radiative efficiency ($\epsilon\equiv\frac{L}{\dot{M}c^2}$). The
radiative efficiency tightly correlates with the accretion rate and
the parameter $\delta$ that describes the fraction of the direct
viscous heating to electrons. Xie \& Yuan (2012) indicated that the
radiative efficiency of hot accretion flows increases with the
accretion rate and is highly enhanced with larger $\delta$.
According to the results of Xie \& Yuan (2012), we couple the
luminosity and the inflow mass rate through the inner boundary of
computational domain in the following way. We set the
`circularization'  radius ($r_{\text{cir}}$) of the environment gas
to be smaller than our inner boundary. Bu et al. (2013) show that
when the gas is beyond the `circularization'  radius,the gas can
freely fall down and no outflow can be produced. Therefore, we
assume that gas goes into the inner boundary of computational domain
will freely fall until the `circularization' radius . Also, we
assume that the mass accretion rate is constant with radius between
the inner boundary and the `circularization' radius. When gas
arrives the `circularization' radius, the gas will continue fall
on to the black hole accompanied by angular momentum transfer by
viscosity.  For hot accretion flows, with high angular momentum, the
mass inflow rate $\dot{M}_{\text{inflow}}\propto r^s$, where
$s\approx0$ for $r\lesssim 10 r_{\text{s}}$ and $s\approx0.5$ for
$10r_s<r<r_{\text{cir}}$ (Yuan et al. 2012; Bu et al. 2013,
2016a,b; Yuan et al. 2015). Thus, we calculate the net accretion rate
$\dot{M}_{\text{net}}$ falling on to the central black hole in the
following formula:
\begin{equation}
\dot{M}_{\text{net}}=\dot{M}_{\text{in}}
(\frac{10r_{\text{s}}}{r_{\text{cir}}})^{0.5},
\end{equation}
where $\dot{M}_{\text{in}}$ is the mass inflow rate thought the
inner boundary of computational domain. Xie \& Yuan (2012)
calculated the radiative efficiency of hot accretion flows for the
case of viscous parameter $\alpha=0.1$ and use a piecewise power-law
function to fit the radiative efficiency for ADAF and type 1 LHAF.
The radiative efficiency is given by
\begin{equation}
\epsilon(\dot{M}_{\text{net}})=\epsilon_0(\frac{100\dot{M}_{\text{net}}}{\dot{M}_{\text{Edd}}})^a,
\end{equation}
where $\epsilon_0$ and $a$ are given in table 1 of Xie \& Yuan (2012) for different $\delta$. We choose the case of $\delta=0.5$ and have

\begin{equation}
(\epsilon_0,a) = \left\{ \begin{array}{ll}
(1.58,0.65) & \textrm{if } \frac{\dot{M}_{\text{net}}}{\dot{M}_{\text{Edd}}}\lesssim2.9\times10^{-5};\\
(0.055,0.076) & \textrm{if } 2.9\times10^{-5}<\frac{\dot{M}_{\text{net}}}{\dot{M}_{\text{Edd}}}\lesssim3.3\times10^{-3};\\
(0.17,1.12) & \textrm{if } 3.3\times10^{-3}<
\frac{\dot{M}_{\text{net}}}{\dot{M}_{\text{Edd}}}\lesssim5.3\times10^{-3}.
\end{array} \right.
\end{equation}
When
$\frac{\dot{M}_{\text{net}}}{\dot{M}_{\text{Edd}}}>5.3\times10^{-3}$,
the radiative efficiency $\epsilon(\dot{M}_{\text{net}})$ is simply
set to be 0.1.

Because of a time lag of accretion flows travelling from the inner
boundary of computation domain to black hole, we calculate the
accretion luminosity at a given time by $L(t)=\epsilon
\overline{\dot{M}}_{\text{net}}(t) c^2$.
$\overline{\dot{M}}_{\text{net}}(t)$ is the average net mass
accretion rate during a time interval $\Delta t$ and we can write it
as $\overline{\dot{M}}_{\text{net}}(t)=\int^{t-\tau}_{t-\tau-\Delta
t}\dot{M}_{\text{net}}(t^{'})dt^{'}/\int^{t-\tau}_{t-\tau-\Delta
t}dt^{'}$, where $\tau$ is the lag time and can be approximated as
the accretion time-scale in the `standard' $\alpha$ disc models. For
simplicity, we set $\Delta t=\tau$. A similar treatment was employed by Kurosawa \& Proga (2009) to determine the feedback luminosity of cool accretion flows.

When the luminosity of a black hole is higher than 2$\% L_{\rm Edd}$,
its spectrum may transit from hard to soft state (e.g. Yuan \& Li
2011). It is hard to determine exact transition luminosity. For
simplicity, 2$\% L_{\rm Edd}$ is taken as switch luminosity between hard
and soft states and as the maximum luminosity of hot accretion
flows.

\subsection{Simulation set-up}

All of our models are assumed to be axisymmetric and calculated in spherical coordinates ($r$,$\theta$,$\phi$). The origin is set at the central black hole of $M=10^8M_{\odot}$ ($M_{\odot}$ is solar mass). The size of the computational domain is $500 r_{\text{s}} \leq r \leq 10^6 r_{\text{s}}$ and $0 \leq \theta \leq \pi/2$. Non-uniform grid is employed in the $r$ direction. The 140 zones with the zone size ratio, $dr_{i+1}/dr_i=1.05$, are distributed in the $r$ direction. The 88 zones are uniformly distributed in the
$\theta$ direction, i.e. $d\theta_{i+1}/d\theta_i=1$.

Initially, the computational domain is filled by gas with uniform density $\rho=\rho_0$ and temperature $T=T_0$. The radial and $\theta$ direction velocities ($v_r, v_{\theta}$) of the gas are zero. The angular velocity ($v_{\phi}$) of the gas is assigned to have the following specific angular momentum distribution $l(\theta)=l_0 (1-|cos(\theta)|)$, where $l_0$ is the specific angular momentum at $r_{\text{cir}}$ at the equatorial plane.

We apply the following boundary conditions. Axis-of-symmetry and reflecting boundary conditions are applied at the pole (i.e.
$\theta=0$) and the equatorial plane (i.e. $\theta=\pi/2$), respectively. The outflow boundary condition is adopted at the inner radial boundary, i.e. all HD variables in the ghost zones are set to the values at the inner radial boundary ($r_{\text{in}}$). At the outer radial boundary ($r_{\text{out}}$), all HD variables except the radial velocity are set to be equal to the initially chosen values, i.e. $\rho=\rho_0$, $T=T_0$, $v_{\theta}=0$, and $v_{\phi}=l/(r sin\theta)$, when $v_r(r_{\text{out}},\theta)<0)$ (inflowing); all HD variables in the ghost zones are set to the values at $r_{\text{out}}$, when $v_r(r_{\text{out}},\theta)>0$ (outflowing).

\section{Results}

We summarize all the models with different parameters in Table 1. As shown in Table 1, we have performed runs 1a$-$14a to examine the effect of changing the outer boundary conditions on accretion luminosity. Our attention is paid to two important parameters, i.e. density ($\rho_0$) and temperature ($T_0$) at the outer boundary. Virial temperature at the outer boundary equals $\sim 10^7 \text{K}$. We consider four different temperatures: $T_0=2\times10^7$, $4\times10^7$, $6\times10^7$ and $10^8$K. The corresponding Bondi radius is located within the computational domain and much less than the outer boundary. For comparison, we have also performed runs 1b$-$3b and 1c$-$3c to examine the roles of circularization radius $r_{\text{cir}}$ and the radiation temperature $T_{\text{X}}$, respectively.

\subsection{Comparison to Bondi accretion}
The gases are spherically symmetrically accreted from a non-rotating polytropic gas cloud to the central BH, which is called Bondi accretion (Bondi 1952). When the adiabatic index $\gamma=5/3$, the Bondi accretion rate is given by
\begin{equation}
\dot{M}_{\rm B} =\pi \rho_{\infty}\frac{G^2 M^2}{c_{\infty}^3},
\end{equation}
where $\rho_{\infty}$ and $c_{\infty}$ are the density and sound speed of gas cloud at infinity, respectively. The Bondi accretion formula predicts $\dot{M}_{\rm B}\varpropto\rho_{\infty}/T_{\infty}^{\frac{3}{2}}$, where $T_{\infty}$ is the gas temperature at infinity. It is noted that the Bondi model ignores magnetic and radiation fields and assumes that the accreted gases are non-rational, inviscid and adiabatic. When these physical factors are considered, the accretion rate significantly deviates from $\dot{M}_{\rm B}$. Proga \& Begelman (2003) and Narayan \& Fabian (2011) find that the accretion rate of slightly rotating gas can be significantly reduced compared to the Bondi accretion rate.

\begin{figure*}
\scalebox{0.5}[0.5]{{\includegraphics[bb=26 5 504 380]{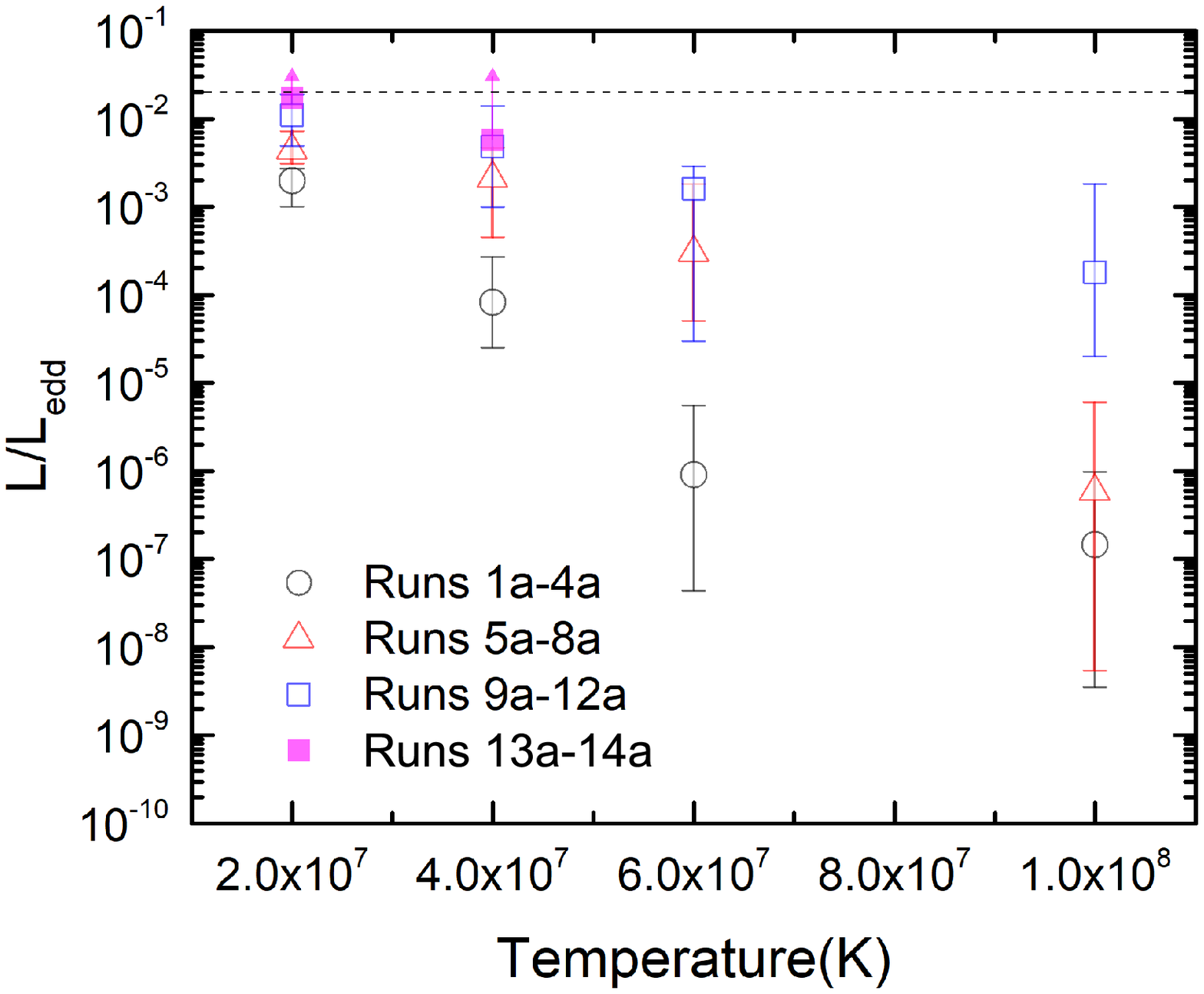}}}
\scalebox{0.5}[0.5]{{\includegraphics[bb=26 5 504 380]{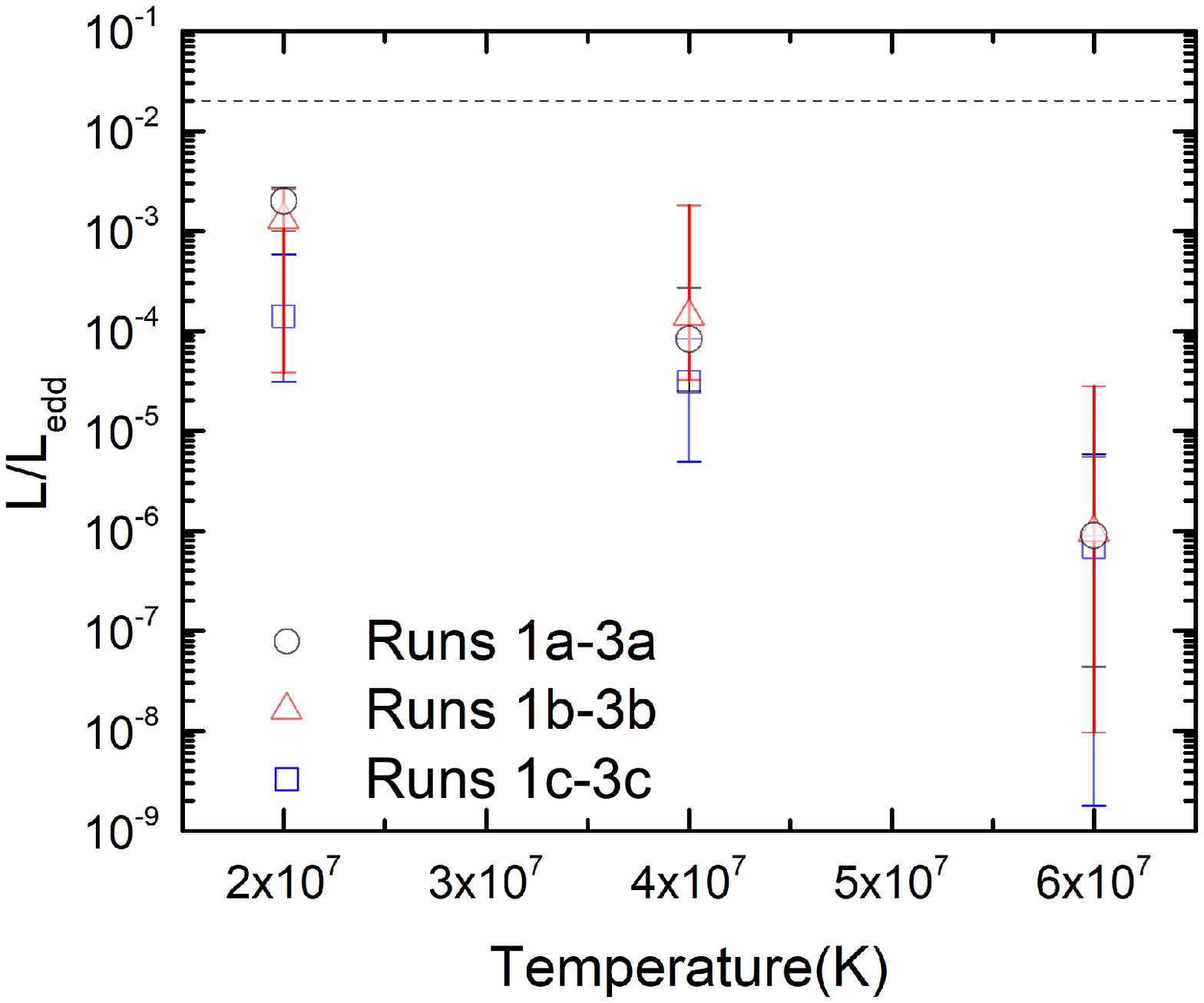}}}
\ \centering \caption{The luminosities of models against the temperatures at the outer boundary. The left-hand panel is for runs 1a$-$14a and the right-hand panel is for runs 1a$-$3a, runs 1b$-$3b and runs 1c$-$3c. Symbols indicate the average values of luminosities and error bars indicate the change range of luminosities. } \label{fig 4}
\end{figure*}

The radial range of computational domain is $5\times10^{-3} \text{ pc}\leq r \leq 9.3 \text{ pc}$, so that the density and temperature at the outer boundary are approximately regarded as the $\rho_{\infty}$ and $T_{\infty}$ of gas at infinity, i.e. $\rho_{\infty}=\rho_0$ and $T_{\infty}=T_0$. For our models, their Bondi radius ($R_{\rm B}\equiv\frac{GM}{c_{\infty}^2}$), inside which the gravitational energy dominates over the thermal energy of the gas, is between $0.5 \text{ pc}$ and $2.6 \text{ pc}$. Because of that the radiation heating and cooling are considered in our models, the simulation results deviate from the Bondi solution, as shown in Fig. 1. Fig. 1 shows the deviation of the mass inflow rate ($\dot{M}_{\rm in}$) (through the inner boundary) from the Bondi accretion rate in our models. From Fig. 1, we can see that the $\dot{M}_{\rm in}$ of our models is either higher or lower than $\dot{M}_{\rm B}$. For example, the $\dot{M}_{\rm in}$ of runs 10a and 11a is one order of magnitude higher than their $\dot{M}_{\rm B}$, while the $\dot{M}_{\rm in}$ of runs 3a, 4a, and 8a is one order of magnitude lower than their $\dot{M}_{\rm B}$ .

Figs 2 and 3 compares the radial profiles of density and temperature in runs 1a, 4a, 9a, and 12a with the Bondi solution. The radial profiles are obtained by the time- and angle-average approach. The $\rho_{0}$ of Run 9a and 12a is three times higher than that of Run 1a and 4a. For all the models, the gas temperature is found to be always comparable to or higher than $10^7 \text{ K}$, so that Compton heating/cooling and bremsstrahlung cooling become the dominant heating and cooling processes. In runs 1a$-$12a, Compton temperature is set to be $10^8 {\rm K}$, so that Compton heating takes place within the domain where the gas temperature is lower than $10^8 {\rm K}$. For example, Compton heating can take place outside $\sim0.2 \text{ pc}$ for runs 1a and 9a, as shown in Figs 2 and 3. For run 1a, within the Bondi radius, the density is significantly increased and the temperature is reduced compared to the Bondi solution. For run 4a, its temperature is comparable to or slightly lower than that of the Bondi solution within the Bondi radius, while its density is lower than that of the Bondi solution. Because of the low density in run 4a, radiation cooling is not important within the Bondi radius for run 4a. For runs 9a and 12a, their density is significantly increased compared to the Bondi solution within the Bondi radius and their temperature is lower than that of the Bondi solution due to radiation cooling.

Therefore, after considering the radiation heating and cooling, the hot accretion flows at large radii can significantly deviate from the Bondi accretion.

\subsection{Effect of Model Parameters on Luminosity}
Fig. 4 shows the luminosity of models as a function of temperatures at the outer boundary. The left-hand panel is for Runs 1a$-$14a and the right-hand panel is for runs 1a$-$3a, runs 1b$-$3b and runs 1c$-$3c, respectively. We can see that the density and temperature of the gas at parsec scale significantly affect the luminosities of hot accretion flows. The luminosity quickly decreases with increasing temperature. Also, luminosity quickly increases with increasing density. For example, in runs 1a$-$4a with $\rho_0=10^{-22} \text{g cm}^{-3}$, when the temperature has been increased from $2\times10^7\text{ K}$ to $10^8\text{ K}$, the luminosity can decrease by four orders of magnitude. This implies that the increase of environment temperatures is unfavourable for accretion processing and the environment temperature at parsec scale is a key parameter to regulate the luminosities of hot accretion flows. The reason for the decrease of luminosity with increasing gas temperature is easy to be understood. With the increase of temperature, the Bondi radius decreases; the mass of gas bound to the central black hole is decreased, which results in a decreasing accretion rate and luminosity.

It is noted that the luminosity of run 13a is never smaller than 2$\% L_{\rm Edd}$. In this work, we take 2$\% L_{\rm Edd}$ as the maximum luminosity of hot accretion flows. The spectrum of a black hole may transit from hard to soft state if its luminosity is higher than 2$\% L_{\rm Edd}$. Here, our simulations indicate that when the environment density at parsec-scale is higher than $4\times10^{-22} \text{g } {\rm cm}^{-3}$, such as run 13a, the spectrum of black hole has a chance to transit from hard to soft state.

The right-hand panel in Fig. 4 shows that runs 1b$-$3b with $r_{\text{cir}}=150 r_{\text{s}}$ have comparable luminosities with runs 1a$-$3a with $r_{\text{cir}}=350 r_{\text{s}}$. The $r_{\text{cir}}$ here is smaller than the inner boundary of computational domain. In fact, the `circularization' radius does not exist for a steady hot accretion flow and the angular momentum profile should be smooth throughout the flow (Bu \& Yuan 2014). In spite of this, we also use the `circularization' radius to determine how much the gas angular momentum is. However, the angular momentum of gas plays a minor role in our models.

For runs 1c$-$3c, we set $T_{\text{X}}=10^9 \text{ K}$. Clearly, with the increase of $T_{\text{X}}$, the luminosity of hot accretion flows decreases. Compared with Run 1a, the luminosity of Run 1c significantly decreases. Because of the large difference between the radiation temperature and the gas temperature, the Compton heating rate is enhanced in run 1c and the irradiated gas easily escapes from the gravity of black hole. The radiation temperature plays a minor role in run 3c because of its higher gas temperature.

\section{Conclusions and Discussions}

We have performed a set of axisymmetric hydrodynamical simulations to study the effects of accretion environment at parsec scale on hot accretion flows. The gas properties at parsec scale can significantly influence the black hole activity. Also, the gas at parsec scale can be affected by the irradiation from the hot accretion flow. Our simulations were performed within the range of $500 r_{\text{s}}-10^6 r_{\text{s}}$. We set $2\% L_{\text{Edd}}$ as the maximum luminosity of hot accretion flow. The luminosity of black hole in this paper is self-consistently determined according to the mass inflow rate through the inner boundary. We take into account Compton heating (cooling), bremsstrahlung cooling, photoionization heating, and line and recombination continuum cooling simultaneously. It is found that the slowly rotating flows within $\sim5\times10^{-3} \text{ pc}\leq r \leq \sim9.3 \text{ pc}$ can significantly deviate from Bondi accretion when radiation heating and cooling are considered. It is further found that the density and temperature of the gas at large radius ($\sim 10 $pc) significantly influence the luminosity of hot accretion flows. When the environment temperature of accretion is relatively low ($T=2\times 10^7$ K), the luminosity of black holes is relatively high. When the environment temperature is relatively high ($T\ga 4\times 10^7$ K), the luminosity decreases significantly. When environment density at large radii is higher than $\sim 4\times10^{-22} \text{g } \text{cm}^{-3}$, the luminosity will always achieve the switch luminosity ($L\ga 2\%L_{\rm Edd}$) from hard to soft state. This implies that the accretion environment with density higher than $\sim 4\times 10^{-22} \text{g } \text{cm}^{-3}$ is not suitable for harbouring a hot accretion flow with $L\ga 2\%L_{\rm Edd}$.

Our simulations considered only the radiation feedback effect from hot accretion flows due to Compton scattering. However, both simulations (e.g.McKinney, Tchekhovskoy \& Blandford 2012; Narayan et al. 2012; Yuan et al. 2012, 2015; Bu et al.2016a,b) and observations (e.g.Tombesi et al. 2010, 2014; Crenshaw \& Kraemer 2012) show that strong wind exists in the hot accretion flows. Yuan et al. (2015) pointed out that the mass flux of wind is very significant for the hot accretion flows and the mass flux of wind is roughly equal to the inflow rate. The wind feedback effects on the accretion environment may be significant and need to be taken into account in future.

We set the `circularization' radius of the gas at large radii to be less than the inner boundary of computational domain. When the `circularization' radius equals 350r$_{\rm s}$, the special angular momentum of gas is $\sim 40\text{ km s}^{-1} \text{pc}$, which is much less than the specific angular momentum of the stellar disc in early-type galaxies (ETGs) and late-type galaxies (LTGs). For the ETGs, the specific angular momentum of the stellar disc is higher than $10^{4.5}\text{ km s}^{-1} \text{pc}$ (Fall 1983; Shi et al. 2017). In future, it is necessary to study the case with large angular momentum. When the angular momentum of the gas is large, the falling time can be much larger than the case in this paper. Also, viscous heating will be present and affect the properties of the gas. In future, we will study the effects of environment gas with large angular momentum on the black hole activity.

\section*{Acknowledgments}
We thank Feng Yuan and Zhao-Ming Gan for useful discussions. This work is supported by the Fundamental Research Funds for the Central Universities (106112016CDJXY300007), the Natural Science Foundation of China (grants
11773053, 11573051, 11633006) and the Natural Science Foundation of
Shanghai (grant 16ZR1442200).


\begin{thebibliography}{99}
\bibitem[\protect\citeauthoryear{Abramowicz et al.}{1988}]{Abramowicz et al. 1988}Abramowicz M. A., Czerny B., Lasota J. P., Szuszkiewicz, E. 1988, ApJ, 332, 646
\bibitem[\protect\citeauthoryear{Bu et al.}{2014}]{Bu et al. 2014)}Bu D. F., Yuan F., Wu M. C., Cuadra J., 2013, MNRAS, 434, 1692
\bibitem[\protect\citeauthoryear{Bu \& Yuan}{2014}]{Bu and Yuan 2014}Bu D. F., \& Yuan F. 2014, MNRAS, 442, 917
\bibitem[\protect\citeauthoryear{Bu et al.}{2016a}]{Bu et al. 2016a}Bu D. F., Yuan F., Gan Z. M., Yang X. H. 2016a, ApJ, 818, 83
\bibitem[\protect\citeauthoryear{Bu et al.}{2016b}]{Bu et al. 2016b}Bu D. F., Yuan F., Gan Z. M., Yang X. H. 2016b, ApJ, 823, 90
\bibitem[\protect\citeauthoryear{Bondi}{1952}]{Bondi 1952}Bondi H. 1952, MNRAS, 112, 195
\bibitem[\protect\citeauthoryear{Ciotti \& Ostriker}{1997}]{Ciotti and Ostriker 1997}Ciotti L., \& Ostriker J. P. 1997, ApJ, 487, L105
\bibitem[\protect\citeauthoryear{Ciotti \& Ostriker}{2001}]{Ciotti and Ostriker 2001}Ciotti L., \& Ostriker J. P. 2001, ApJ, 551, 131
\bibitem[\protect\citeauthoryear{Ciotti \& Ostriker}{2007}]{Ciotti abd Ostriker 2007} Ciotti L., \& Ostriker J. P. 2007, ApJ, 665, 1038
\bibitem[\protect\citeauthoryear{Ciotti, Ostriker \& Proga}{2009}]{Ciotti, Ostriker and Proga 2009} Ciotti L., Ostriker J. P., \& Proga D. 2009, ApJ, 699, 89
\bibitem[\protect\citeauthoryear{Ciotti et al.}{2017}]{Ciotti et al. 2017} Ciotti L., Pellegrini, S., Negri , A., Ostriker J. P. 2017, ApJ, 835, 15
\bibitem[\protect\citeauthoryear{Crenshaw \& Kraemer}{2012}]{Crenshaw and Kraemer 2012} Crenshaw D. M, Kraemer S. B. 2012, ApJ, 753, 75
\bibitem[\protect\citeauthoryear{Di Matteo et al.}{2005}]{Di Matteo et al. 2005} Di Matteo T., Springel V., Hernquist L. 2005, Nature, 433, 604
\bibitem[\protect\citeauthoryear{Fall}{1983}]{Fall 1983} Fall, S. M. 1983, in IAU Symposium, Vol. 100, Internal
Kinematics and Dynamics of Galaxies, ed. E. Athanassoula, 391¨C398
\bibitem[\protect\citeauthoryear{Ferrarese \& Merritt}{2000}]{Ferrarese and Merritt 2000} Ferrarese L., Merritt D. 2000, ApJ, 539, L9
\bibitem[\protect\citeauthoryear{Gan et al.}{2014}]{Gan et al. 2014} Gan Z. M., Yuan F., Ostriker J. P., Ciotti L., Novak G. S. 2014, ApJ, 789, 150
\bibitem[\protect\citeauthoryear{Gebhardt}{2000}]{Gebhardt 2000} Gebhardt K. et al. 2000, ApJ, 539, L13
\bibitem[\protect\citeauthoryear{Khandai et al.}{2015}]{Khandai et al. 2015} Khandai N., Di Matteo T., Croft R., et al. 2015, MNRAS, 450, 1349
\bibitem[\protect\citeauthoryear{Kurosawa \& Proga}{2009}]{Kurosawa and Proga 2009} Kurosawa R., \& Proga D. 2009, MNRAS, 397, 1791
\bibitem[\protect\citeauthoryear{Kormendy \& Bender}{2009}]{Kormendy and Bender 2009} Kormendy J., \& Bender R., 2009, ApJ, 691, L142
\bibitem[\protect\citeauthoryear{Hayes et al.}{2006}]{Hayes et al. 2006} Hayes J. c., Norman M. L., Fiedler R. A., Border J. O., et al. 2006, ApJS, 165, 188
\bibitem[\protect\citeauthoryear{Magorrian et al.}{1998}]{Magorrian et al. 1998} Magorrian J., et al. 1998, AJ, 115, 2285
\bibitem[\protect\citeauthoryear{Mckinney et al.}{2012}]{Mckinney et al. 2012} Mckinney J. C., Tchekhovskoy A., \& Blandford R. D. 2012, MNRAS, 423, 3083
\bibitem[\protect\citeauthoryear{Narayan \& Yi}{1994}]{Narayan and Yi 1994} Narayan R., \& Yi I. 1994, ApJ, 428, L13
\bibitem[\protect\citeauthoryear{Narayan \& Yi}{1995}]{Narayan and Yi 1995} Narayan R., \& Yi I. 1995, ApJ, 452, 710
\bibitem[\protect\citeauthoryear{Narayan \& McClintock}{2008}]{Narayan and McClintock 2008} Narayan R., \& McClintock J. E. 2008, NewAR, 51, 733
\bibitem[\protect\citeauthoryear{Narayan et al.}{2012}]{Narayan et al. 2012} Narayan R., Sadowski A., Penna R. F., Kulkarni A. K. 2012, MNRAS, 426, 3241
\bibitem[\protect\citeauthoryear{Narayan \& Fabian}{2011}]{Narayan and Fabian 2011} Narayan, R., Fabian, A. C. 2011, MNRAS, 415, 3721
\bibitem[\protect\citeauthoryear{Negri \& Volonteri}{2017}]{Negri and Volonteri 2017} Negri A., \& Volonteri M. 2017, MNRAS, 467, 3475
\bibitem[\protect\citeauthoryear{Ostriker et al.}{2010}]{Ostriker et al. 2010} Ostriker J. P., Choi E., Ciotti L., Novak G. S., Proga D. 2010, ApJ, 722, 642
\bibitem[\protect\citeauthoryear{Paczy\'{n}sky \& Wiita}{1980}]{Paczynsky and Wiita 1980} Paczy\'{n}sky, B., \& Wiita, P. J. 1980, A\&A, 88, 23
\bibitem[\protect\citeauthoryear{Pellegrini}{2005}]{Pellegrini 2005} Pellegrini, S. 2005, ApJ, 624, 155
\bibitem[\protect\citeauthoryear{Proga et al.}{2000}]{Proga et al. 2000} Proga D., Stone J. M., \& Kallman T. R. 2000, ApJ, 543, 686
\bibitem[\protect\citeauthoryear{Proga et al.}{2003}]{Proga et al. 2003} Proga D., \& Begelman M. C. 2003, ApJ, 582, 69
\bibitem[\protect\citeauthoryear{Proga}{2007}]{Proga 2007} Proga D. 2007, ApJ, 661, 693
\bibitem[\protect\citeauthoryear{Proga et al.}{2008}]{Proga et al. 2008} Proga D., Ostriker J. P., Kurosawa R. 2008, ApJ, 676, 101
\bibitem[\protect\citeauthoryear{Liu et al.}{2013}]{Liu et al. 2013} Liu C., Yuan F., Ostriker J. P., Gan Z. M., Yang X. H. 2013, MNRAS, 434, 1721
\bibitem[\protect\citeauthoryear{Sazonov et al.}{2005}]{Sazonov et al. 2005} Sazonov S. Y., Ostriker J. P., Ciotti L., \& Sunyaev, R. A. 2005, MNRAS, 358, 168
\bibitem[\protect\citeauthoryear{Springel et al.}{2005}]{Springel et al. 2005} Springel V., Di Matteo T., Hernquist L. 2005, MNRAS, 361, 776
\bibitem[\protect\citeauthoryear{Shakura \& Sunyaev}{1973}]{Shakura and Sunyaev 1973} Shakura N. I., \& Sunyaev R. A. 1973, A\&A, 24, 337
\bibitem[\protect\citeauthoryear{shi et al.}{2017}]{Shi 2017} Shi, J., Lapi, A., Mancuso, C., et al. 2017, ApJ, 843, 105
\bibitem[\protect\citeauthoryear{Soria et al.}{2006}]{Sorie et al. 2006} Soria, R. Fabbiano, G., Graham, A. W., et al. 2006, ApJ, 640, 126
\bibitem[\protect\citeauthoryear{Teyssier et al.}{2011}]{Teyssier et al. 2011} Teyssier R., Moore B., Martizzi D., et al. 2011, MNRAS, 414, 195
\bibitem[\protect\citeauthoryear{Tombesi et al.}{2010}]{Tombesi et al. 2010} Tombesi F., Sambruna R. M., Beeves J. N., et al. 2010, ApJ, 719, 700
\bibitem[\protect\citeauthoryear{Tombesi et al.}{2014}]{Tombesi et al. 2014} Tombesi F., Tazaki F., Mushotzky R. F., et al. 2014, MNRAS, 443, 2154
\bibitem[\protect\citeauthoryear{Vogelsberger et al.}{2013}]{Vogelsberger et al. 2013} Vogelsberger M., Genel S., Sijacki D., et al. 2014, MNRAS, 444, 1518
\bibitem[\protect\citeauthoryear{Xie \& Yuan}{2012}]{Xie and Yuan 2012}Xie F. G., \& Yuan F. 2012, MNRAS, 427, 1580
\bibitem[\protect\citeauthoryear{Xie et al.}{2017}]{Xie et al. 2017} Xie F. G., Yuan F., Ho L. C. 2017, ApJ, 844, 42
\bibitem[\protect\citeauthoryear{Yuan}{2001}]{Yuan 2001} Yuan F., 2001, MNRAS, 324, 119
\bibitem[\protect\citeauthoryear{Yuan}{2003}]{Yuan 2003} Yuan F., 2003, ApJ, 594, L99
\bibitem[\protect\citeauthoryear{Yuan \& Xie}{2011}]{Yuan and Xie 2011}Yuan F., Xie F. Ostriker J. P. 2009, ApJ, 691, 98
\bibitem[\protect\citeauthoryear{Yuan \& Li}{2011}]{Yuan and Li 2011}Yuan F., \& Li M. 2011, ApJ, 737, 23
\bibitem[\protect\citeauthoryear{Yuan, Bu \& Wu}{2012}]{Yuan, Bu and Wu 2012} Yuan F., Bu D. F., \& Wu M. C.  2012, ApJ, 761, 130
\bibitem[\protect\citeauthoryear{Yuan \& Narayan}{2014}]{Yuan and Narayan 2014} Yuan F., \& Narayan, R.  2014, ARA\&A, 52, 529
\bibitem[\protect\citeauthoryear{Yuan et al.}{2015}]{Yuan et al. 2015} Yuan F., Gan Z. M., Narayan, R., et al.  2015, ApJ, 804, 101


\end{thebibliography}
\end{document}